\documentstyle[preprint,aps,floats,epsf]{revtex}
\tightenlines
\begin{document}

\newcommand{\notE}{\ \hbox{{$E$}\kern-.60em\hbox{/}}}
\newcommand{\notp}{\ \hbox{{$p$}\kern-.43em\hbox{/}}}


\preprint{
\font\fortssbx=cmssbx10 scaled \magstep2
\hbox to \hsize{
\hbox{\fortssbx University of Wisconsin - Madison}
\hfill$\vcenter{\hbox{\bf MADPH-98-1028}
                \hbox{\bf hep-ph/9804451}
                \hbox{April 1998}}$ }
}
 
\title{\vspace*{2cm}
Trilepton Signal of Minimal Supergravity \\
at the Tevatron Including $\tau$-lepton Contributions}
 
\author{V. Barger, Chung Kao and Tianjun Li}

\address{
Department of Physics, University of Wisconsin, 
Madison, WI 53706, USA}

\maketitle

\thispagestyle{empty}
 
\bigskip

\begin{abstract}

The trilepton signal with missing transverse energy 
($3\ell+\notE_T$, with $\ell = e$ or $\mu$) 
from chargino-neutralino ($\chi^\pm_1 \chi^0_2$) associated production 
and decays is studied for the upgraded Fermilab Tevatron Collider with 2 TeV 
center of mass energy and integrated luminosity of 2 fb$^{-1}$ (MI) 
to 20 fb$^{-1}$ (TeV33). 
In some regions of parameter space in the minimal supergravity model, 
$\chi^\pm_1$ and $\chi^0_2$ decay dominantly into final states 
with $\tau$ leptons via real or virtual $\tilde{\tau}_1$ sleptons.  
The contributions from $\tau-$leptonic decays increase the trilepton signal 
from $\chi^\pm_1 \chi^0_2$ by at least a factor of two 
when soft but realistic cuts on lepton transverse momenta are used.
With the Main Injector, a trilepton signal can be detected 
at $\tan\beta \equiv v_2/v_1 \sim 3$  
for universal masses $m_{1/2}, m_0 \alt$ 200 GeV.

\end{abstract}

%


\section{Introduction}


In the near future, the Main Injector (MI) of the Fermilab Tevatron 
will run at 2 TeV center of mass energy 
and accumulate an integrated luminosity ($L$) of about 2 fb$^{-1}$ 
at each of the CDF and the D$\emptyset$ detectors. 
It has been proposed to further upgrade the Tevatron luminosity to 
$10^{33}$ cm$^{-2}$ s$^{-1}$ (TeV33) to obtain an integrated luminosity 
$L = 20$ fb$^{-1}$ \cite{Gillman}. 
Such a great increase in luminosity will significantly improve the potential 
of Tevatron to discover new physics beyond the Standard Model (SM) 
before the CERN Large Hadron Collider (LHC) begins operation \cite{TeV2000}. 

In this article, we assess the prospects for discovery 
of the trilepton signal of supersymmetry at the upgraded Tevatron. 
The source of this signal is associated production of 
the lightest chargino ($\chi^\pm_1$) 
and the second lightest neutralino ($\chi^0_2$) with decays to leptons 
\cite{Trilepton1,Trilepton2}. 
For definiteness, we base our analysis on supersymmetric particle masses 
and couplings of the minimal supergravity (mSUGRA) model \cite{SUGRA}. 
In the minimal supergravity model with gauge coupling unification, 
the sleptons ($\tilde{\ell}$), the lighter chargino ($\chi^\pm_1$) 
and the lighter neutralinos ($\chi^0_1,\chi^0_2$) 
are considerably less massive than the gluinos and squarks 
over most of the parameter space. 
Because of this, 
the trilepton signal with missing transverse energy ($3\ell+\notE_T$) 
from associated production of the lightest chargino ($\chi^\pm_1$) 
and the second lightest neutralino ($\chi^0_2$) 
\cite{Trilepton1,Trilepton2,CDF,D0} 
is one of the most promising channels for supersymmetric particle searches 
at the Tevatron.  
The background to this signal from SM processes can be greatly reduced 
with suitable cuts.
A principal departure of our analysis from previous studies is the inclusion 
of the contributions from $\tau$-lepton decays to leptons 
through the use of softer transverse momentum ($p_T$) acceptance cuts 
on leptons. 
The softer cuts increase the observable signal by at least a factor of two 
over results with conventional hard cuts \cite{Trilepton2}. 
Our soft cuts are similar to those of a recent CDF analysis \cite{CDF98} 
and should be realistic at the upgraded Tevatron as well \cite{Demina&Kamon}. 


A supersymmetry (SUSY) between fermions and bosons provides 
a natural explanation of the Higgs mechanism 
for electroweak symmetry breaking (EWSB) 
in the framework of a grand unified theory (GUT). 
For the particle content 
of the minimal supersymmetric standard model (MSSM) \cite{MSSM}, 
the evolution of gauge couplings 
by renormalization group equations (RGEs) \cite{RGE} 
is consistent with a grand unified scale 
at $M_{\rm GUT} \sim 2\times 10^{16}$ GeV 
and an effective SUSY mass scale in the range 
$M_Z < M_{\rm SUSY} \alt 10$ TeV \cite{Unification}. 
With a large top quark Yukawa coupling ($Y_t$) to a Higgs boson 
at the GUT scale, radiative corrections drive 
the corresponding Higgs boson mass squared parameter negative, 
spontaneously breaking the electroweak symmetry 
and naturally explaining the origin of the electroweak scale.
In the minimal supersymmetric GUT with a large $Y_t$, 
there is a quasi-infrared fixed point (IRFP) \cite{BBO,IRFP} 
at low $\tan\beta$; the top quark mass is correspondingly predicted 
to be $m_t = (200 \; {\rm GeV}) \sin\beta$ \cite{BBO}, 
and thus $\tan\beta \simeq 1.8$ for $m_t = 175$ GeV. 
At high $\tan\beta$, another IRFP solution ($dY_t/dt \simeq 0$) exists 
at $\tan\beta \sim 56$.


In supergravity (SUGRA) models \cite{SUGRA}, 
supersymmetry is broken in a hidden sector 
with SUSY breaking communicated to the observable sector 
through gravitational interactions, 
leading naturally but not necessarily \cite{Non-universal} 
to a common scalar mass ($m_0$), a common gaugino mass ($m_{1/2}$), 
a common trilinear coupling ($A_0$) and a common bilinear coupling ($B_0$) 
at the GUT scale.
Through minimization of the Higgs potential, 
the $B$ coupling parameter of the superpotential  
and the magnitude of the Higgs mixing parameter $\mu$ are related to 
the ratio of Higgs-field vacuum expectation values (VEVs) 
($\tan\beta \equiv v_2/v_1$) and to the mass of the $Z$ boson ($M_Z$).
The SUSY particle masses and couplings at the weak scale 
can be predicted by the evolution of RGEs 
from the unification scale \cite{BBO,SUGRA2}. 
In most of the mSUGRA parameter space, 
the weak-scale gaugino masses are related 
to the universal gaugino mass parameter $m_{1/2}$ by 
$m_{\chi^0_1} \sim 0.44 m_{1/2}$ and 
$m_{\chi^\pm_1} \sim m_{\chi^0_2} \sim 0.88 m_{1/2}$. 
Consequently, this discovery channel could provide valuable information 
about the value of $m_{1/2}$.


Recent measurements of the $b \to s\gamma$ decay rate 
by the CLEO \cite{CLEO} and LEP collaborations \cite{LEP} 
place constraints on the parameter space 
of the minimal supergravity model \cite{bsg}.
It was found that $b \to s\gamma$ excludes 
most of the minimal supergravity (mSUGRA) parameter space 
when $\tan\beta$ is large and $\mu > 0$ \cite{bsg}. 
Therefore, we will choose $\mu < 0$ in our analysis since we are particularly 
interested in large $\tan\beta$.
In our convention, 
$-\mu$ appears in the chargino mass matrix and 
$+\mu$ appears in the neutralino mass matrix. 


The Yukawa couplings of the bottom quark ($b$) and the tau lepton ($\tau$) 
are proportional to $\tan\beta$ and are thus greatly enhanced 
when $\tan\beta$ is large.
In SUSY GUTS, the masses of the third generation sfermions 
are consequently very sensitive to the value of $\tan\beta$.
As $\tan\beta$ increases, the lighter tau slepton ($\tilde{\tau}_1$) 
and the lighter bottom squark ($\tilde{b}_1$) become lighter than 
charginos and neutralinos while other sleptons and squarks are heavy.
Then, $\chi^\pm_1$ and $\chi^0_2$ can dominantly decay into 
final states with tau leptons via real or virtual $\tilde{\tau}_1$. 
While these decays reduce the trilepton signal with hard cuts 
to suppress the backgrounds, 
they also open new discovery channels via $\tau$ leptons. 

One way to detect $\tau$ leptons is through their one prong 
and three prong hadronic decays. 
The CDF and the D$\emptyset$ collaborations are currently investigating 
the efficiencies for detecting these modes 
and for implementing a $\tau$ trigger. 
Recently, it has been suggested that the $\tau$ leptons 
in the final state may be a promising way 
to search for $\chi^\pm_1 \chi^0_2$ production at the Tevatron 
if excellent $\tau$ identification becomes feasible \cite{BCDPT,Wells}.

Another way of exploiting the $\tau$ signals, 
that we consider in this article, 
is to detect the soft electrons and muons from leptonic $\tau$ decays 
by employing softer but realistic $p_T$ cuts on the leptons 
than conventionally used. 
We find that this can improve the trilepton signal 
from $\chi^\pm_1 \chi^0_2$ production by at least a factor of two.


The relevance of $\tau$ leptons in the production 
and decays of $\chi^\pm_1 \chi^0_2$, are illustrated in Figure 1,
where the product of the cross section 
$\sigma(p\bar{p} \to \chi^\pm_1 \chi^0_2 +X)$ 
and the branching fraction 
$B(\chi^\pm_1 \chi^0_2 \to 3 \; leptons +\notE)$ versus $\tan\beta$ 
is presented with $\mu < 0$, $m_{1/2} = 200$ GeV, $m_0 =$ 100 and 200 GeV, 
for four final states (a) $\tau\tau\tau$, (b) $\tau\tau \ell$, 
(c) $\tau\tau\tau$ and (d) $\ell\ell\ell$, where $\ell = e$ or $\mu$.
For $m_0 \alt$ 200 GeV and/or $\tan\beta \agt 40$, 
channels with at least one $\tau$ lepton are dominant.


\section{Production Cross Section and Branching Fractions}


In hadron collisions, associated production of chargino and neutralino 
occurs via quark-antiquark annihilation 
in the $s$-channel through a virtual $W$ boson 
($q\bar{q'} \to {W^{\pm}}^* \to \chi^\pm_1 \chi^0_2$) and 
in the $t$ and $u$-channels through squark ($\tilde{q}$) exchanges.
The $p\bar{p} \to \chi^\pm_1 \chi^0_2 +X$ cross section 
depends mainly on the masses of chargino ($m_{\chi^\pm_1}$) and 
neutralino ($m_{\chi^0_2}$).
For squarks much heavier than the gauginos, 
the $s$-channel $W$-resonance amplitude dominates.
If the squarks are light, 
a destructive interference between the $W$ boson and the squark exchange 
amplitudes can suppress the cross section by as much as $40\%$ 
compared to the $s$-channel contribution alone.
For larger squark masses, 
the effect of negative interference is reduced, and 
the cross section is enhanced.


In Figure 2, we present branching fractions of $\chi^0_2$ and $\chi^\pm_1$ 
versus $\tan\beta$ for $\mu < 0$, $m_{1/2} = 200$ GeV 
and several values of $m_0$.
For $\tan\beta \alt 5$, 
the branching fractions are sensitive to the sign of $\mu$.

For $\mu < 0$ and $\tan\beta \sim 3$, we observe that: 
\begin{itemize}
\item For $m_0 \alt 50$ GeV, 
$\chi^0_2$ decays dominantly to 
$\tilde{\nu}_L\nu, \tilde{\ell}_R\ell$ and $\tilde{\tau}_1\tau$, 
and $\chi^\pm_1$ decays into $\tilde{\nu}_L\ell$ and $\tilde{\tau}_1\nu$, 
\item For 60 GeV $\alt m_0 \alt$ 110 GeV, 
the $\chi^0_2 \to \tilde{\nu}_L\nu$ decay is kinematically suppressed; 
$\chi^0_2$ decays mainly to $\tilde{\ell}_R \ell$ and $\tilde{\tau}_1\tau$, 
and $\chi^\pm_1$ decays dominantly into $\tilde{\tau}_1\nu$. 
\item For 120 GeV $\alt m_0 \alt$ 170 GeV,   
all two-body $\chi$ decays to sleptons are closed.
However, 
the $\chi^\pm_1 \chi^0_2 \to 3 \ell +\notE$ braching fractions  
is still significant due to relatively light virtual sleptons.
\item For $m_0 \agt$ 180 GeV, 
$\chi^\pm_1$ and $\chi^0_2$ dominantly decay into  $q\bar{q}'\chi^0_1$.
\end{itemize}

For $\mu > 0$ and $\tan\beta \sim 3$, we observe that: 
\begin{itemize}
\item For $m_0 \alt$ 100 GeV, 
$\tilde{\ell}_R$, $\tilde{\ell}_L$ and $\tilde{\nu}_L$ are all lighter than 
$\chi^\pm_1$ and $\chi^0_2$; 
$\chi^0_2$ dominantly decays to $\tilde{\nu}_L\nu$
and $\chi^\pm_1$ decays into $\tilde{\nu}_L\ell$.
\item For 110 GeV $\alt m_0 \alt$ 140 GeV, 
$\tilde{\ell}_R$, and $\tilde{\tau}_1$ are lighter than 
$\chi^\pm_1$ and $\chi^0_2$; 
$\chi^0_2$ dominantly decays to $\tilde{\tau}_1\tau$
and $\chi^\pm_1$ decays into $\tilde{\tau}_1\nu$.
\item For 140 GeV $\alt m_0 \alt$ 160 GeV,   
all two-body $\chi$ decays to sleptons are closed.
However, the $\chi^\pm_1 \chi^0_2 \to 3 \ell +\notE$ branching fraction 
is still significant due to relatively light virtual sleptons.
\item For $m_0 \agt$ 170 GeV, 
$\chi^\pm_1$ and $\chi^0_2$ dominantly decay into  $q\bar{q}'\chi^0_1$.
\end{itemize}

For $\mu < 0$ and $m_0 \sim $ 200 GeV, $\chi^0_2$ dominantly decays 
(i) into $\tau\bar{\tau}\chi^0_1$ for $25 \alt \tan\beta \alt 40$, 
(ii) into $b\bar{b}\chi^0_1$ for $\tan\beta \simeq 40$, and
(iii) into $\tau\bar{\tau}_1$ for $\tan\beta \agt 40$.  
For $m_0 \alt 300$ GeV and large $\tan\beta \agt 35$,
both $\tilde{\tau}_1$ and $\tilde{b}_1$ can be lighter than other sfermions, 
and $\chi^\pm_1$ and $\chi^0_2$ can decay dominantly into final states 
with $\tau$ leptons or $b$ quarks 
via virtual or real $\tilde{\tau}_1$ and $\tilde{b}_1$. 
For $m_0 \agt 400$ GeV and $4 \alt \tan\beta \alt 40$, 
$B(\chi^0_2 \to \tau^+\tau^- \chi^0_1) 
\sim B(\chi^0_2 \to e^+e^- \chi^0_1) \sim 2\%$

\section{Discovery Potential at the Tevatron}

In this section, we present results from simulations 
with an event generator and a simple calorimeter 
including our acceptance cuts. 
The ISAJET 7.37 event generator program \cite{ISAJET} is employed 
to calculate the $3\ell +\notE_T$ signal from $\chi^\pm_1 \chi^0_2$ 
and the dominant backgrounds from $t\bar{t}$ and $WZ$ 
at the upgraded Tevatron. 
A calorimeter with segmentation 
$\Delta\eta \times \Delta\phi = 0.1 \times 0.1$ 
extending to $|\eta| = 4$ is used. 
An energy resolution of $\frac{0.7}{\sqrt{E}}$ for the hadronic calorimeter 
and $\frac{0.15}{\sqrt{E}}$ for the electromagnetic calorimeter is assumed. 
Jets are defined to be hadron clusters with $E_T > 15$ GeV in a cone 
with $\Delta R=\sqrt{\Delta\eta^2+\Delta\phi^2}=0.7$. 
Leptons with $p_T > 5$ GeV and within $|\eta_{\ell}| < 2.5$ 
are considered to be isolated if the hadronic
scalar $E_T$ in a cone with $\Delta R = 0.4$ about the lepton is smaller
than 2 GeV.

The cuts we implement are very similar to those of recently employed CDF cuts 
\cite{CDF98}. \\
(i) We require 3 isolated leptons in each event, with 
\begin{eqnarray}
p_T(\ell_1)    & > & 12 \; {\rm GeV}, \;\; 
p_T(\ell_2,\ell_3) >  5 \; {\rm GeV}, \;\; {\rm and} \nonumber \\
|\eta(\ell_1)| & < & 1.0, \;\; |\eta(\ell_2,\ell_3)| < 2.4. 
\end{eqnarray}
(ii) We require $\notE_T > 25$ GeV. 
This cut removes backgrounds from SM processes such as Drell-Yan dilepton
production, where an accompanying jet may fake a lepton. \\
(iii) To reduce the background from $WZ$ production, 
we require that the invariant mass of any opposite-sign dilepton pair 
not reconstruct the $Z$ mass: $|m(\ell\bar{\ell})-M_Z|\geq 10$ GeV. 
This cut reduces $WZ$ background to below the 1 fb level. \\
The surviving total background cross section from $WZ$ and $t\bar{t}$ 
is about 0.47 fb. 
The background is mainly due to $WZ$ events, where $Z\to\tau\bar{\tau}$, 
with subsequent $\tau$ lepton decays. 

The effect of cuts on the signal and background is demonstrated in Table I.
The cross sections of the signal with $m_{1/2} = 200$ GeV, $m_0 = 100$ GeV, 
and several values of $\tan\beta$, 
along with $WZ$ and $t\bar{t}$ backgrounds 
are presented for three cases: (a) No Cuts, (b) Soft Cuts, (c) Hard Cuts.
We note that the soft cuts enhance the signal 
by factors of two to eight for $3 \alt \tan\beta \alt 25$, 
while the backgrounds only increase by about $50\%$.

At the upgraded Tevatron with the Main Injector 
and 2 fb$^{-1}$ of integrated luminosity, 
we expect about one background event from the background cross section 
of 0.47 fb.
The signal cross section is required to yield 
a minimum of four signal events for discovery. 
The Poisson probability for the SM background to fluctuate 
to this level is less than $0.4\%$.
For TeV33 with $L = 20$ fb$^{-1}$, we expect about 9 events of background 
and a $5 \sigma$ signal would be 15 events, 
which corresponds to a signal cross section of 0.77 fb. 

The cross sections of the trilepton signal and background after cuts 
are shown versus $m_0$ in Figure 3 for $m_{1/2} = 200$ GeV 
and $\tan\beta =$ 1.8, 10 and 35.
With $L = 2$ fb$^{-1}$ and $m_{1/2} =$ 200 GeV, 
the trilepton signal may be observable 
for $m_0 \alt 350$ GeV with $\tan\beta \sim 1.8$ and 
for $m_0 \alt 150$ GeV with $\tan\beta \sim 10$.
With $L = 20$ fb$^{-1}$ and $m_{1/2} =$ 200 GeV, 
the trilepton signal may be observable 
for $m_0 \alt 550$ GeV with $\tan\beta \sim 1.8$ and 
for $m_0 \agt 500$ GeV with $\tan\beta \agt 10$.
For 180 GeV $\alt m_0 \alt$ 400 GeV and $10 \alt \tan\beta \alt 40$, 
the $\chi^0_2$ decays dominantly into $q\bar{q}\chi^0_1$ and 
in these regions it will be difficult to establish a supersymmetry signal.

To further assess the discovery potential for the upgraded Tevatron, 
we present the cross sections of the trilepton signal and background 
after cuts versus $m_{1/2}$ in Figure 4 for $\mu < 0$, $\tan\beta =$ 3 and 35
and $m_0 =$ 100, 200, 500 and 1000 GeV.
Also shown are the curves for 
(i) 4 signal events with $L = 2$ fb$^{-1}$ and 
(ii) a 5$\sigma$ signal with $L = 20$ fb$^{-1}$. 
The reach in $m_{1/2}$ at the MI and the TeV33 is presented 
in Tables II and III.

\section{Conclusions}

In some regions of the mSUGRA parameter space, 
the $\chi^\pm_1$ and the $\chi^0_2$ decay 
dominantly to final states with $\tau$ leptons. 
The subsequent leptonic decays of these $\tau$ leptons 
contribute importantly to the trilepton signal
from $\chi^\pm_1 \chi^0_2$ associated production.
With soft but realistic lepton $p_T$ acceptance cuts, 
these $\tau \to \ell$ contributions enhance the trilepton signal 
by at least a factor of two. 
The branching fractions of $\chi^\pm_1$ and $\chi^0_2$ decays 
into $\tau$ leptons are dominant when the universal scalar mass $m_0$
is less than about 200 GeV and/or $\tan\beta \agt 40$.

The Tevatron trilepton searches are most sensitive to the region 
of mSUGRA parameter space with $m_0 \alt 100$ GeV and $\tan\beta \alt$ 10.
\begin{itemize}
\item For $m_0 \sim 100$ GeV and $\tan\beta\sim 3$, 
the trilepton signal should be detectable 
(i) at the MI if $m_{1/2} \alt$ 260 GeV, and
(ii) at the TeV33 if $m_{1/2} \alt$ 290 GeV.
\item For $m_0 \sim 150$ GeV and $\tan\beta\sim 35$, 
the trilepton signal should be detectable 
at the TeV33 if $m_{1/2} \alt$ 170 GeV.
\item For $m_0 \agt 500$ GeV and $\tan\beta\sim 35$, 
the trilepton signal should be detectable 
at the TeV33 if $m_{1/2} \alt$ 200 GeV.
\end{itemize}
A difficult region for the trilepton search at the upgraded Tevatron 
is 180 GeV $\alt m_0 \alt$ 400 GeV with $10 \alt \tan\beta \alt 35$, 
because for these parameters $\chi^\pm_1$ and $\chi^0_2$ 
dominantly decay into  $q\bar{q}'\chi^0_1$.


\section*{Acknowledgments}

We are grateful to Howie Baer, Chih-Hao Chen, Xerxes Tata 
and Dieter Zeppenfeld for useful discussions 
and to Frank Paige for instructions about running ISAJET 
on Digital UNIX ALPHA workstations.
V.B. and C.K. would like to thank the Aspen Center for Physics 
where part of the research was completed. 
This research was supported in part by the U.S. Department of Energy 
under Grants No. DE-FG02-95ER40896 
and in part by the University of Wisconsin Research Committee
with funds granted by the Wisconsin Alumni Research Foundation.


%

\newpage


\medskip

\begin{table}
\caption{
The effect of cuts on the trilepton signal 
(with $m_{1/2} = 200$ GeV, $m_0 = 100$ GeV) 
and the background from $WZ$ and $t\bar{t}$ 
at the upgraded Tevatron:
(a) No Cuts;  
(b) Soft Cuts: 
$p_T(\ell_1) >$ 12 GeV, $\eta(\ell_1)| <$ 1.0,
$p_T(\ell_2,\ell_3) >$ 5 GeV, $\eta(\ell_2,\ell_3)| <$ 2.4, 
$|E_T(cone) < 2|$ GeV, and $|M_{\ell\ell} - M_Z| > 10$ GeV;
(c) Hard Cuts: 
$p_T(\ell_1) >$ 20 GeV, $p_T(\ell_2) >$ 15 GeV, 
$p_T(\ell_3) >$ 10 GeV, $|\eta(\ell's)| <$  2.5, 
$|E_T(cone) < p_T(\ell)/4|$, and $|M_{\ell\ell} - M_Z| > 10$ GeV.
}
\medskip
\begin{tabular}{lccccccc} 
Case $\backslash \tan\beta$ & 3 & 10 & 15 & 20 & 25 & $WZ$ & $t\bar{t}$ \\
\tableline
No Cuts   & 40  & 14  & 8.9  & 7.4  & 6.9    & 37   &  - \\
Soft Cuts & 12  & 2.6 & 1.1  & 0.67 & 0.21   & 0.40 &  0.073 \\
Hard Cuts & 6.1 & 1.2 & 0.44 & 0.16 & 0.0028 & 0.28 &  0.041
\end{tabular}
\end{table}


\bigskip

\begin{table}
\caption{
The anticipated $m_{1/2}$ reach (in GeV) from a trilepton search 
at the upgraded Tevatron with $L = 2$ fb$^{-1}$ (MI) 
for various values of $\tan\beta$ and $m_0$.
Note that, for $m_0 < 400$ GeV, 
the trilepton cross section is too small for discovery, 
especially if $10 < \tan\beta < 40$.
}
\medskip
\begin{tabular}{lcccc} 
$\tan\beta$ & $m_0({\rm GeV})=$ 100 & 200 & 500 & 1000 \\
\tableline
 3 & $m_{1/2}({\rm GeV})=$ 260 & 200 & 130 & 150 \\
35 & $m_{1/2}({\rm GeV})=$ $ < 100$   & $ < 100$ & 160 & 190 
\end{tabular}
\end{table}


\bigskip

\begin{table}
\caption{
The anticipated $m_{1/2}$ reach (in GeV) from a trilepton search 
at the upgraded Tevatron with $L = 20$ fb$^{-1}$ (TeV33) 
for various values of $\tan\beta$ and $m_0$.
}
\medskip
\begin{tabular}{lcccc} 
$\tan\beta$ & $m_0({\rm GeV})=$ 100 & 200 & 500 & 1000 \\
\tableline
 3 & $m_{1/2}({\rm GeV})=$ 290 & 230 & 170 & 200 \\
35 & $m_{1/2}({\rm GeV})=$ 160 & 110 & 200 & 230
\end{tabular}
\end{table}


\begin{figure}
\centering\leavevmode
\epsfxsize=6in\epsffile{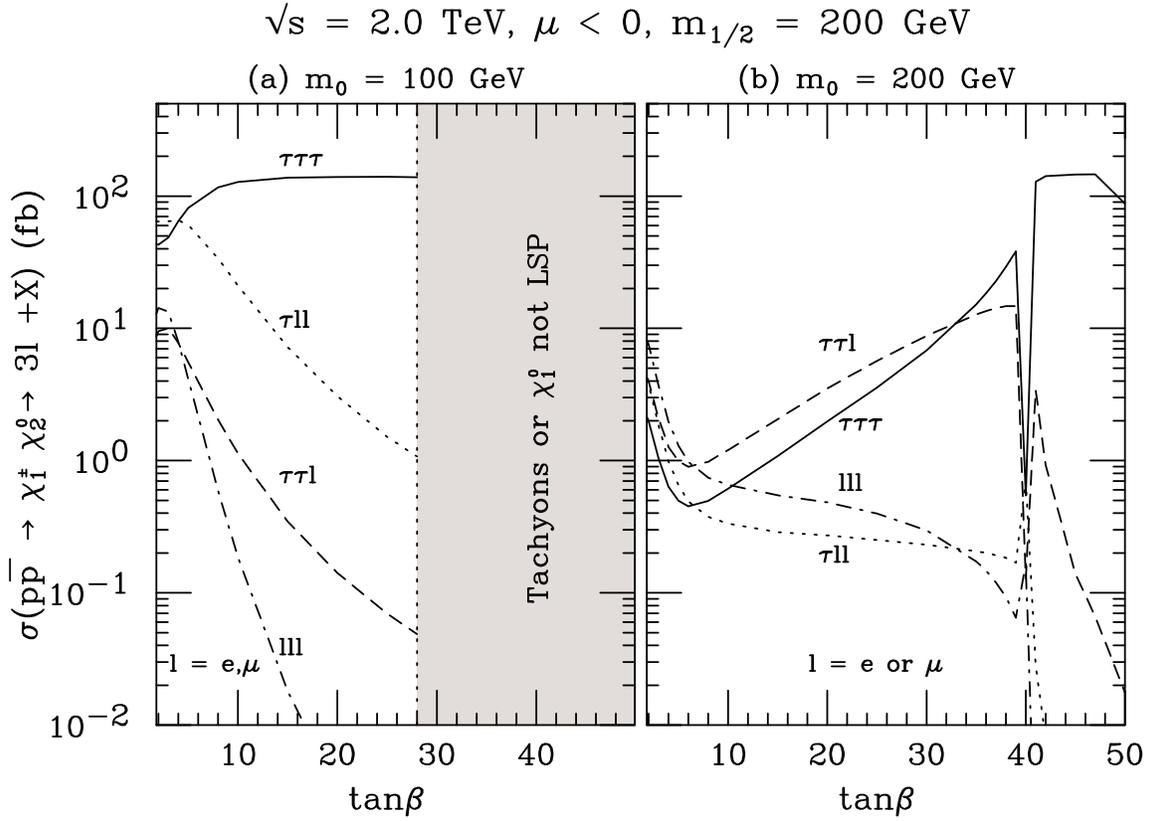}

\bigskip
\caption[]{
Cross section of $p\bar{p} \to \chi^\pm_1 \chi^0_2 \to 3\ell's +X$ 
without cuts at $\sqrt{s} = 2$ TeV verses $\tan\beta$, 
with $\mu < 0$, $m_{1/2} = 200$ GeV, $m_0 =$ 100 GeV 
for 4 final states: (a) $\tau\tau\tau$ (solid), 
(b) $\tau \tau \ell$ (dot-dash), 
(c) $\tau \ell \ell$ (dash) and (d) $\ell \ell \ell$ (dot), 
where $\ell = e$ or $\mu$. 
\label{fig:taus}
}\end{figure}

\begin{figure}
\centering\leavevmode
\epsfxsize=4.75in\epsffile{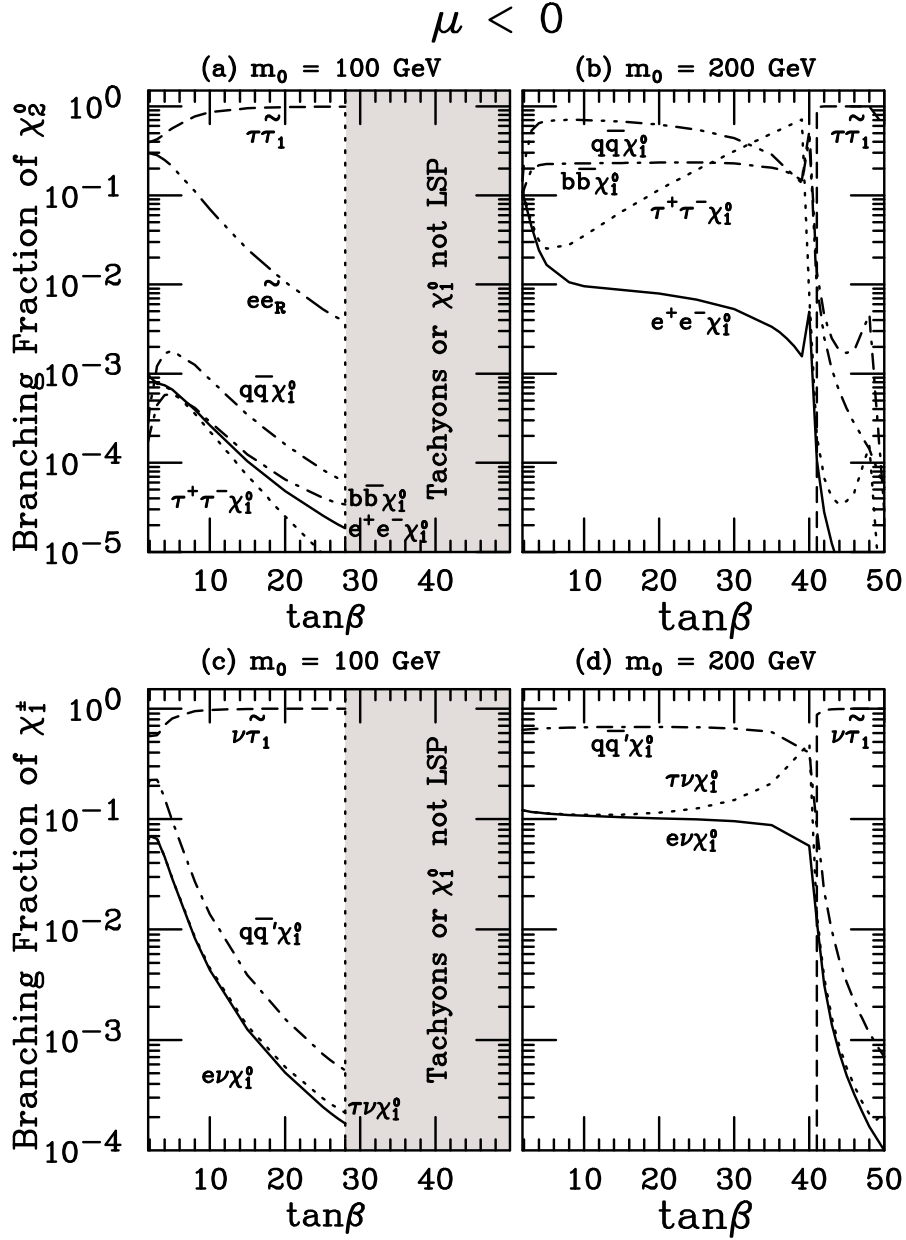}

\bigskip
\caption[]{
Branching fractions of $\chi^0_2$ and $\chi^\pm_1$ 
decays into various channels versus $\tan\beta$ with $m_{1/2} =$ 200 GeV, 
for $m_0 = 100$ GeV [(a) and (c)] as well as $m_0 = 200$ GeV [(b) and (d)].
\label{fig:bfwz}
}\end{figure}

\begin{figure}
\centering\leavevmode
\epsfxsize=6in\epsffile{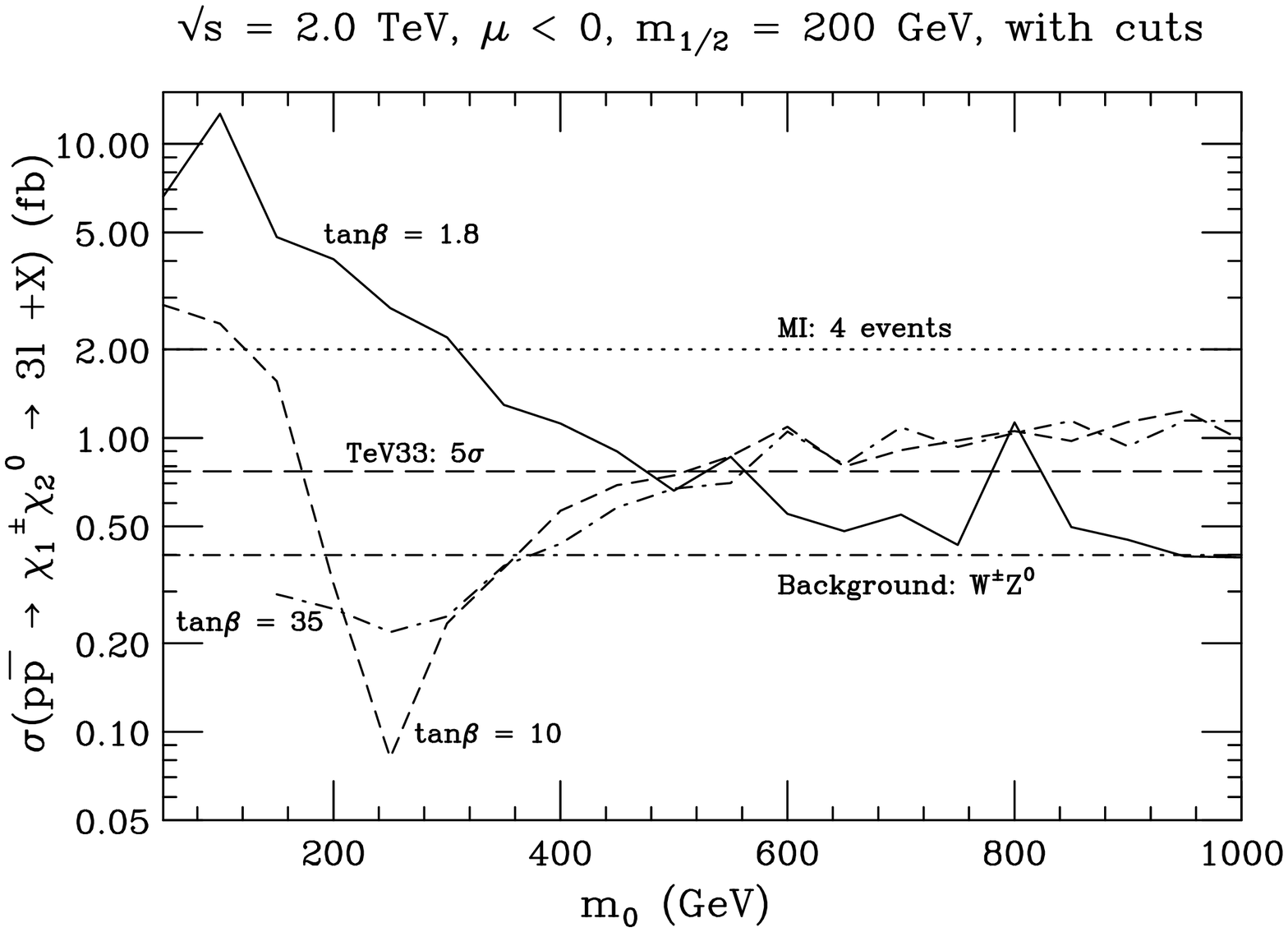}

\bigskip
\caption[]{
Cross section of $p\bar{p} \to \chi^\pm_1 \chi^0_2 \to 3\ell +X$ 
at $\sqrt{s} = 2$ TeV versus $m_0$, with soft cuts (Eq.~[1]), 
for $m_{1/2} = 200$ GeV, $\mu < 0$ 
and $\tan\beta =$ 1.8 (solid), 10 (dash) and 35 (dot-dash).
Also noted is the cross section of background from $WZ$ (dot-dash), 
for 5 events with $L =$ 2 fb$^{-1}$ (dot) and 
for 5 $\sigma$ with $L =$ 20 fb$^{-1}$ (dash).
\label{fig:xm0}
}\end{figure}

\begin{figure}
\centering\leavevmode
\epsfxsize=6in\epsffile{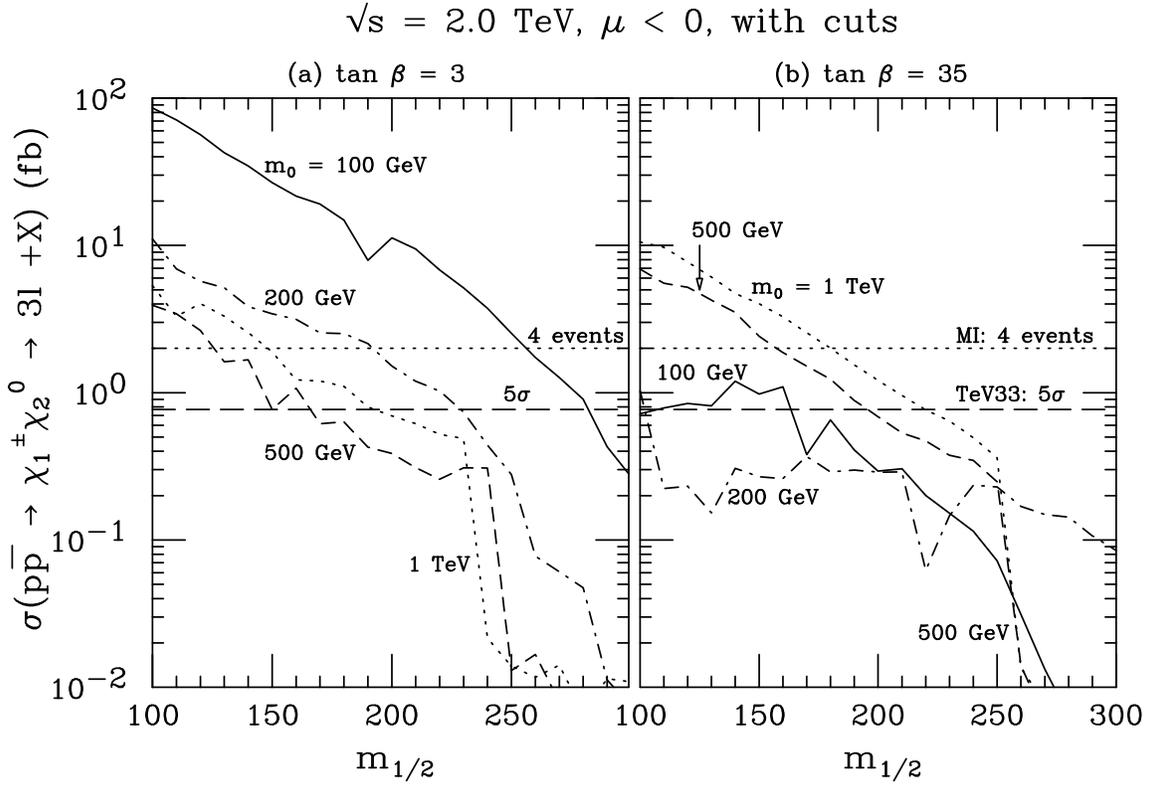}

\bigskip
\caption[]{
Cross section of $p\bar{p} \to \chi^\pm_1 \chi^0_2 \to 3\ell +X$ 
at $\sqrt{s} = 2$ TeV, with soft cuts versus $m_{1/2}$, 
with (a) $\tan\beta =$ 3 and (b) $\tan\beta = 35$, 
for $m_0 =$ 100 GeV (solid), 200 GeV (dot-dash), 500 GeV (dash) 
and 1000 GeV (dot).
Also noted is the cross section for 5 events with $L =$ 2 fb$^{-1}$ (dot) and 
5 $\sigma$ for $L =$ 20 fb$^{-1}$ (dash).
\label{fig:xmhf}
}\end{figure}
%

\end{document}